\documentclass[a4paper]{article}

\usepackage[english]{babel}
\usepackage[utf8x]{inputenc}
\usepackage[T1]{fontenc}

\usepackage[a4paper,top=3cm,bottom=2cm,left=3cm,right=3cm,marginparwidth=1.75cm]{geometry}

\usepackage{amssymb}
\usepackage{amsmath}
\usepackage{graphicx}
\usepackage[colorinlistoftodos]{todonotes}
\usepackage[colorlinks=true, allcolors=blue]{hyperref}

\title{Observing positive and negative AGN feedback}
\author{Giovanni Cresci and Roberto Maiolino}

\begin{document}
\maketitle


\textbf{Galaxy-scale outflows powered by actively accreting supermassive black holes are routinely detected, and they have been associated both with suppression and triggering of star formation. Recent observational evidence and simulations are favouring a delayed mechanism that connects outflows and star formation.}\\


Massive and fast outflows are almost ubiquitous in luminous active galaxies. 
These outflows are detected in different gas phases and physical scales, from pc-scale ultra-fast ($\sim 10\%$ of the speed of light) outflows detected in X-rays to kpc-scale outflows detected in atomic, molecular and ionized gas, with velocities up to $\sim1000$\ km\ s$^{-1}$. 
Several models have suggested that these massive and fast outflows can suppress star formation in the host galaxy, by removing and heating ISM. However, from the observational point of view it has not yet been clearly assessed if and how the presence of spatially-unresolved outflows relates to the level of star formation in the host galaxies\cite{Woo17}
. Studies have in fact found that statistically there is no relationship between the two, although this may not be unexpected if the impact of the outflows is delayed with respect to the timescale of visible AGN activity \cite{Harrison16}.

The advent of Integral Field Units (IFU) has allowed us to spatially resolve the ionized outflows, and to start studying their impact on the physical properties of the host galaxies. 
Near-IR IFU observations of $z\sim1-3$ quasars have revealed a spatial anti-correlation between the location of the fast outflowing gas (as traced by the blue-shifted wing of the [OIII] line]) and the star formation in the host galaxy (as traced by the decomposed narrow component of H$\alpha$ least affected by the AGN\cite{canodiaz12,cresci15a,carniani16}.
In all these objects the outflow appears to affect the gas reservoir only along its path, while  star formation remains globally high, with star-formation rates (SFRs) as high as $\approx100$\ M$_{\odot}$\ yr$^{-1}$. 
However, one should take into account that H$\alpha$ is subject to dust extinction, hence these results should be confirmed with extinction-free star formation tracers (such as far-IR maps).

An alternative approach to evaluate the impact of feedback is the use of CO observations at mm wavelengths to probe the molecular gas content of AGN hosts. The derived gas fractions can be compared with those of galaxies with similar mass (and SFR), to probe gas depletion potentially associated with gas ejection in AGN with outflows when compared with the population of normal star forming galaxies. 
However, this approach is complicated by several factors, such as: (1) the difficulty in separating AGN and host-galaxy emission to derive stellar masses; (2) the need for high-quality far infrared observations to reliably derive SFRs; (3) the uncertainty in conversion factors between CO luminosities and molecular gas masses; (4) defining a well matched non-AGN control sample. 
Nevertheless, when this exercise has been attempted for a few obscured AGN, lower gas fractions and shorter gas depletion timescales in targets with AGN-driven outflows compared to star forming galaxies of similar mass and SFR have been reported \cite{brusa17}.


Despite the limitations discussed above and the small number of sources for which direct evidence of feedback in action has been found, the available data suggest that a significant amount of gas is entrained in AGN driven outflows. However, only a small fraction of the outflowing gas may escape the host halo, while a large fraction may fall back onto the galaxy at later times \cite{arribas14}. Furthermore, even for the few sources showing evidence of feedback, the overall impact does not seem to result into a global shut down of star formation\cite{cresci15a}. Consequently, the data do not seem to favour a purely ejective feedback with immediate quenching. These observations agree with some recent high-resolution simulations, where AGN-driven outflows may only affect star-formation in the central regions, even in the most extreme quasars \cite{Costa17}. 
As mentioned above, a delayed feedback mechanism, 
where star formation is more slowly fading across the whole disk with a time scale of roughly $\approx1$ Gyr, would be more consistent with observations. In support to this scenario,In support to this scenario,the comparison of stellar metallicity in star forming and passive galaxies suggests that gas consumption by star formation without a replenishment of fresh gas is the primary mechanism for quenching star formation \cite{peng15}, rather than rapid gas ejection. 
Such delayed feedback may result from the outflow injecting energy into the halo, heating the gas and preventing it from cooling, so that accretion of the fresh supply of gas needed to fuel star formation is halted (``preventive feedback''\cite{Pillepich17}). 

\begin{figure}
\centering
\includegraphics[width=1.0\textwidth]{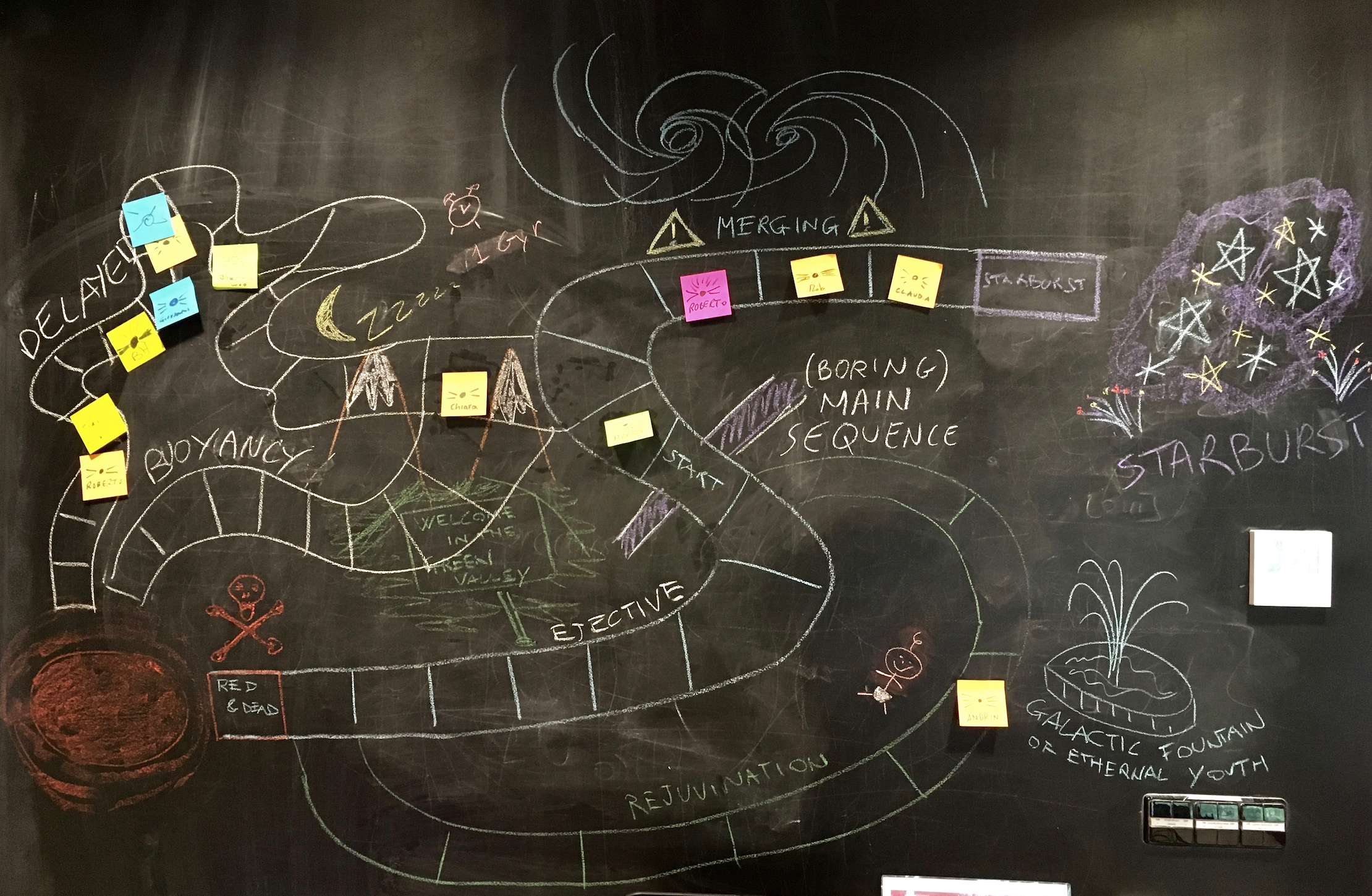}
\caption{\label{fig:blackboard}\textit{The ``Snakes and Ladders'' game played by the participants of the discussion session on negative/positive feedback at the Lorentz Center workshop ``The reality and Myths of AGN feedback'' (Leiden, 16-20 October 2017). Multiple additional paths for quenching, rejuvenation etc.\ were added to the first simple sequence from starburst to red \& dead galaxies. The general consensus reached, as traced by the coloured sticky notes (``ballots'' eventually posted by the participants to the session), was that ``positive feedback'' is present in AGN hosts but its impact on galaxy evolution has yet to be fully assessed, while ``negative feedback'' does not affect the full gas reservoir in an ejective mode, but rather with some form of delayed preventive feedback.}}
\end{figure}

An additional important aspect of fast outflows is that they can also induce star formation, both in the galactic disk through compression of molecular clouds \cite{silk13} or directly in the outflowing gas \cite{ishibashi12}
. These ``positive feedback'' mechanisms have been proposed by several theoretical works to explain observed correlations between AGN luminosity and nuclear SFR, and even bulge formation. Still, very few observations of such phenomenon at work are available, except for those relating the spatial alignment of HII clouds with very powerful radio jets \cite{santoro16}. However, the characterization of this process has been hampered by observational limits and, in particular, by the fact that in AGN-driven outflows the AGN gas excitation generally dominates the diagnostics even if extensive (triggered) star formation is present, and by the difficulties in establishing a causal connection between the ongoing star formation and the AGN activity.
Spatially resolved IFU spectroscopy has partly enabled us to overcome these limitations: evidence for star formation triggering inside the galactic disk has been proposed both in one of the $z\sim1.6$ quasars discussed above \cite{cresci15a}, as well as in the local Seyfert galaxy NGC~5643 \cite{cresci15b}. In both cases, the contribution of positive feedback to the global SFR of the galaxy is $>10\%$, depending on the extinction assumed. Recently, the first evidence of the second mode of positive feedback has also been obtained: star formation has been detected inside the powerful outflow of a nearby ULIRG, with associated SFR\ $>15$\ M$_{\odot}$\ yr$^{-1}$ \cite{maiolino17}. This is supported by the detection of blue-shifted absorption features by young stars in the same outflow region, confirming that the high molecular gas densities in galactic scale outflows provide the appropriate physical condition to form stars in the fast moving gas. 
The need for high spectral and spatial resolution on a broad wavelength range (in order to detect all required diagnostics)  makes the detection of these features of star formation in outflows very challenging. This means that star formation in outflows may be common in galaxies, potentially contributing to the morphological evolution of galaxies as well as to the evolution in size and velocity dispersion of the spheroidal component. Indeed, evidence for several additional examples of star formation in outflows has been reported 
(Gallagher et al., in prep.)\\

The impact of AGN feedback on their host galaxies is still to be firmly assessed. However, with accumulating evidence supporting a \textit{preventive} (i.e. delayed) rather than \textit{ejective} role in quenching star formation, the perception among experts also appears to align with this picture (see Fig.~\ref{fig:blackboard}). Larger samples of galaxies with spatially-resolved observations of outflows and unbiased star formation tracers, especially at the peak of the feedback epoch at $1<z<3$, will be crucial to further test this scenario. The first evidence of AGN-induced star formation is now available, both in the host galaxy and directly in the outflowing gas; given the difficulties in detecting such signatures, dedicated campaigns on large IFU samples are required to start assessing the relevance of this mechanism in the global evolution of galaxies. \\

\textbf{Acknowledgements:}

\textit{This Comment is based on discussion at the Lorentz Center workshop ``The reality and Myths of AGN feedback'', Leiden 16-20 October 2017. We thank all the participants of the workshop for their suggestions, in particular Chris Harrison, Chiara Circosta, Marcella Brusa, Jan Scholtz, Giacomo Venturi, Michele Perna, Giustina Vietri. RM acknowledge support from ERC advanced grant 695671 “QUENCH”}.

\end{document}